%Paper: gr-qc/9504043
%From: "Joshua N. Goldberg" <jngoldbe@mailbox.syr.edu>
%Date: Tue, 25 Apr 1995 17:18:31 -0400 (EDT)
%Date (revised): Fri, 28 Apr 1995 14:23:40 -0400 (EDT)

\def\a{\alpha}

\def\d{\delta}
\def\e{\epsilon}

\def\r{\rho}
\def\s{\sigma}
\def\n{\eta}

\def\o{\omega}
\def\l{\lambda}

\def\ba{{\bf \alpha}}
\def\bb{{\bf \beta}}
\def\bg{{\bf \gamma}}
\def\bd{{\bf \delta}}

\def\bm{{\bf \mu}}
\def\bn{{\bf \nu}}

\def\Ga{\Gamma}

\def\S{\Sigma}

\def\A{{\cal A}}

\def\G{{\cal G}}
\def\H{{\cal H}}

\def\cL{{\cal L}}
\def\M{{\cal M}}

\def\V{{\cal V}}

\def\bi{{\bf i}}
\def\bj{{\bf j}}

\def\0{{\bf 0}}
\def\1{{\bf 1}}
\def\2{{\bf 2}}
\def\3{{\bf 3}}

\def\Sa{\S^{\1}}

\def\Sc{\S^{\3}}
\def\h{\textstyle{1\over2}}

\def\and{\ {\rm and}\ }

\def\rt{\sqrt}
\def\={\ =\ }
\def\pl{\ +\ }
\def\-{\ -\ }
\def\rd{\partial}

\def\Io{1\over}

\def\tv{{\tilde v}}

\def\edth{\partial \llap/}
\def\bedth{\bar\partial \llap/}
\def\ut#1{#1\llap{\lower2ex\hbox{$\widetilde{\hphantom{#1}}$}}}

\def\rtar{\rightarrow}
\def\Rtar{\Rightarrow}

\def\Ab{A^{\2}{}}
\def\Ac{A^{\3}{}}
\def\Sa{\S_{\1}{}}

\def\Sc{\S_{\3}{}}

\medskip
gr-qc/9504043
\centerline {Canonical General Relativity on a Null Surface}
\centerline {with  Coordinate and Gauge Fixing}
\smallskip
\centerline {J.N. Goldberg}
\centerline {Department of Physics, Syracuse University, Syracuse, NY 13244-
1130}
\centerline {and}
\centerline  {C. Soteriou}
\centerline {Department of Mathematics, King's College
London, Strand, London WC2R 2LS}
\medskip
\centerline {ABSTRACT}

We use the canonical formalism developed together with David Robinson to study
the Einstein equations on a null surface.  Coordinate and gauge conditions are
introduced to fix the triad and the coordinates on the null surface.  Together
with the previously found constraints, these form a sufficient number of second
class constraints so that the phase space is reduced to one pair of canonically
conjugate variables:  $\Ac_2\and\Sc^2$.  The formalism is related to both the
Bondi-Sachs and the Newman-Penrose methods of studying the gravitational field
at null infinity.  Asymptotic solutions in the vicinity of null infinity which
exclude logarithmic behavior require the connection to fall off like $1/r^3$
after the Minkowski limit.  This, of course, gives the previous results of
Bondi-Sachs and Newman-Penrose.  Introducing terms which fall off more slowly
leads to logarithmic behavior which leaves null infinity intact, allows for
meaningful gravitational radiation, but the peeling theorem does not extend to
$\Psi_1$ in the terminology of Newman-Penrose.  The conclusions are
in agreement with
those of Chrusciel, MacCallum, and Singleton.
This work was begun as a preliminary study of a reduced phase space
for quantization of general relativity.
\medskip
\noindent {1.}  Introduction.

The canonical approach to quantum gravity received a strong impetus from the
introduction of the new variables by Abhay Ashtekar [1-4].  The use of a
self-dual
connection one-form and a vector density triad as canonical variables leads to
a Hamiltonian which is a polynomial of degree four in the new variables.  This
structure suggests a simplification of the canonical formalism which may lead
to significant easing of the problems of quantization. By now there is a
considerable literature detailing the efforts of many people to understand how
to make effective use of these variables [3,4].

As part of this overall effort, together with David Robinson, we have renewed
the study of canonical general relativity on a null surface [5].
This program had
previously been undertaken using the metric or a tetrad as the configuration
space variables [6,7].  Although these efforts did not recover all the Einstein
equations in a natural way, their principal drawback is that the resulting
system of second class constraints is complicated by the non-polynomial
structure so that there does not appear to be any hope that a successful Dirac
quantization [8,9] can be carried out.  Our hope was that use of the self-dual
formalism on a null surface would retain the polynomial structure of the
Hamiltonian and the constraints.  If so, the second class constraints might not
be as formidable as in the previous treatments.

While some of the second class constraints are indeed simpler, for the most
part they are sufficiently complicated that they cannot easily be eliminated
either directly or by means of the introduction of Dirac brackets [8-10].
Therefore, as a first step toward considering a reduced phase space
quantization, we have repeated the analysis of the gravitational field in the
vicinity of null infinity.  The use of the Ashtekar variables for this analysis
falls between the metric formulation of Bondi [11] and Sachs [12] and the spin
coefficient method of Newman-Penrose [13,14].  Although the full complement
of Dirac
brackets cannot be obtained, a machine caculation did show that the bracket of
one component of the connection (related to the expansion and shear) with its
conjugate triad density is the same as the corresponding Poisson bracket.  This
identified for us which variables should serve as the true dynamical variables
in terms of which to express the remaining components of the gravitational
field.

By focussing attention on the connection, we were immediately struck by how
deep in a ${1/r}$ expansion one must go in order to avoid the appearance of
logarithmic behavior.  Therefore, the question naturally arises whether such
logarithmic behavior is consistent with asymptotic flatness and finite energy.
While we were still analyzing this question, Chrusciel, MacCallum, and
Singleton [15] showed that this is indeed the case.  However, because in this
work we use a different formalism in which the need to examine the logarithmic
terms is glaring, it is worthwhile to present the results.  Also, there is a
difference in our solutions.  We find that even with the appearance of
logarithmic behavior, the coefficient of $1/r$ in the connection must vanish
whereas they do not.  A detailed discussion of our solution leading to this
difference is left for the Appendix.

To see how
far the connection must drop in $1/r$ in order to avoid logarithmic behavior,
we shall first carry
out the analysis requiring an expansion in $1/r$ without logarithms
and then go back to
indicate what changes would be introduced by the logarithmic terms.  The
simplification in the calculations by neglect of logarithms is enormous.
In agreement with the earlier work of Novak and Goldberg [16, 17], the present
results and those of Chrusciel, MacCallum, and Singleton show that null
infinity can be defined and that energy-momentum and radiation of
energy-momentum remain finite.  In our case, logarithmic terms
come in below those responsible for radiation.  In the more general
considerations of Chrusciel et al, the coefficients of the leading logarithmic
terms are independent of the time.  This point does not come up in our work{}.
However, in both studies, the logarithmic terms come in
before those needed to define angular momentum.

In the following section, we shall give a streamlined review of the
construction of the Hamiltonian on a null surface [5].  In section 3, we
introduce our coordinate and tetrad conditions and
present our analysis of the equations, listing the order in which they are to
be solved.  The solution in the absence of
logarithmic terms is given in section 4 and in section 5 we
discuss the logarithmic terms.  We close with a discussion of our results.
\vfill
\eject

\noindent Section 2.  The Hamiltonian.

To obtain the Hamiltonian, we started from a complex Lagrangian
constructed from the self-dual part of the Riemann tensor, following similar
work by Jacobson and Smolin [18] and by Samuel [19] on a space-like surface{}.
We assume the space-time to be real, but consider complex solutions of the
Einstein equations.  After the calculation has been completed, we impose
reality conditions on the variables and recover a real metric and curvature{}.
Because a null surface is degenerate, we lose one of the field equations if we
allow the initial surface to be null from the beginning.  Therefore, in order
to be sure that we recover all of the Einstein equations, we
included an auxiliary variable $\a$.  The surfaces
$t=$constant are space-like, time-like, or null, when $\a < 0, > 0, {\rm or}
= 0$, respectively.   To guarantee that the surfaces would be null, we adjoined
$\a^2=0$ to the Lagrangian with a Lagrange multiplier.  Moreover, because
working on a null surface imposes constraints not present on a space-like
surface, we did not \`a priori eliminate any variables as non-dynamical.
Therefore, we started with a phase space of 40 variables.  The 12 first and
14 second class constraints leave two dynamical degrees of freedom per
hypersurface point as is appropriate on a null surface [20].
An examination of the resulting structure shows that we may limit the
phase space variables to the nine components of the connection and of the
densitized triad vectors on the hypersurface.  We
shall do so below.

We introduce the null basis of one forms and the dual tetrad basis
$$
\eqalign{\theta^{\0}&\=(N+\a\nu^{\1}{}_i N^i)dt+\a\nu^{\1}{}_i dx^i,\cr
\theta^{\bi}&\=\nu^{\bi}{}_i(N^i dt+dx^i)\cr
e_{\0}&\= N^{-1}(\rd_t-N^i\rd_i),\cr
e_{\bi}&\=(v_{\bi}{}^i+\a_{\bi} N^{-1}N^i)\rd_i
-\a_{\bi} N^{-1}\rd_t,\cr}          \eqno(2.1)
$$
where $\a_{\bi}=\a\d^1{}_{\bi}\and \nu^{\bi}{}_i v_{\bj}{}^i\=\d^{\bi}{}_{\bj}$.
All indices have the range 1-3 and repeated indices sum.  Bold face indices
refer to the one forms and tetrads.  The signature is -2,  with
$$
\eqalign {ds^2 & = \theta^{\0}\otimes \theta^{\1} + \theta^{\1} \otimes
\theta^{\0} -
\theta^{\2}\otimes \theta^{\3} -  \theta^{\3}\otimes \theta^{\2},\cr
&= \n_{\ba\bb} \theta^{\ba}\otimes \theta^{\bb}.\cr}
$$
(Greek letters range and sum from zero to three.)
It follows that the surfaces $t=$constant are null surfaces when $\a=0${}.

We choose the orientation so that the volume form is
$$V=-i \theta^{\0} \wedge \theta^{\1} \wedge \theta^{\2} \wedge \theta^{\3}.
 \eqno (2.2)$$
Thus, $-iN\nu$ is
positive, where $\nu$ is the determinent of $\nu^{\bi }{}_i$
and the components of the Levi-Civita tensor with respect
to this basis, $\e_{\ba\bb\bg\bd}$ and
$\epsilon^{\ba\bb\bg\bd}$, satisfy
$$\e_{\0\1\2\3} = \epsilon^{\0\1\2\3} = -i. \eqno (2.3)$$

The connection coefficients are defined by
$$
 d\theta^{\ba} \= \theta^{\bb}\wedge \o^{\ba}{}_{\bb} \eqno (2.4)
$$
and the Riemann tensor by
$$
d\omega^{\ba}{}_{\bb} + \omega^{\ba}{}_{\bg} \wedge\o^{\bg}{}_{\bb} - {1\over2}   R^{\ba}{}_{\bb\bg\bd}\theta^{\bg}\wedge \theta^{\bd}=:
R^{\ba}{}_{\bb}.     \eqno (2.5)
$$

We take as a basis of self-dual two-forms
 $$ \eqalign
{S^{\1}&={1\over2}(\theta^{\1}\wedge\theta^{\0}+\theta^{\3}\wedge\theta^{\2}),\cr
S^{\2} &= \theta^{\1}\wedge \theta^{\2}, \cr
S^{\3} &= \theta^{\3} \wedge \theta^{\0}.\cr} \eqno (2.6) $$
These two-forms satisfy the equations $(A, B= \1, \2, \3)$
 $$ \eqalign {
{}^*\!S^A &\= i S^A \cr
 S^A \wedge S^B &\= i g^{AB} V. \cr} \eqno (2.7)
$$
The $SO(3)$ invariant metric
 $$
g_{AB} = \pmatrix {2  &0 &0 \cr
 0 &0 &-1 \cr
 0 &-1 &0 \cr}\eqno (2.8a)
$$
and its inverse,
$$
g^{AB} = \pmatrix {\h &0 &0 \cr
 0 &0 &-1\cr  0 &-1 &0 \cr} \eqno (2.8b)
$$
are used to raise and lower the uppercase
Latin self-dual, triad indices.

The self-dual components of the connection are
$$
{}^+\!\omega^{\ba\bb} = {1\over 2} ( \omega^{\ba\bb} -
{i\over 2} \epsilon^{\ba\bb}{}_{\bm\bn} \omega^{\bm\bn})
$$
and are represented by $\Ga^A,$:
$$
\Ga^1:=\ {\Io 2} (\o^{\bf {01}}+\o^{\bf {23}}),\quad \Ga^2:=\ \o^{\bf {21}},
\quad \Ga^3:=\ \o^{\bf {03}}.   \eqno (2.9)
$$
{}From these we obtain the self-dual components of the Riemann tensor as
$$
{\Io 2} R^A = d\Ga^a \pl \n^A{}_{BC} \Ga^B\wedge \Ga^C.  \eqno (2.10)
$$

In a $3+1$ decomposition, we have
$$
\Ga^A\= A^A{}_i dx^i + B^A dt, \eqno(2.11)
$$
and
$$ \eqalign {
R^A{}_{ij} & \= 2A^A{}_{[i,j]} \pl 2\n^A{}_{BC}A^B{}_j A^C{}_i, \cr
R^A{}_{0i}  &\= D_i B^A \- A^A{}_{i,0}. \cr } \eqno (2.12)
$$
The derivative operator $D$ acts on the index $A$ as
$$
D_i f^A:=\rd_i f^A+2\n^A{}_{BC}A^B{}_i f^C.
$$

The complex action introduced by Jacobson and Smolin [18] and by Samuel [19] is
$$
 S=\int R^A\wedge S^B g_{AB}. \eqno (2.13)
$$
With the above decomposition, the Lagrangian density formed using the self-dual
curvature tensor takes the form
$$
\eqalign{\cL\=&\bigl(A^A{}_{i,0}\S_A{}^i+B^AD_i\S_A{}^i +
R^A{}_{ij}N^i\S_A{}^j-\cr
&-{\ut N}v^i(R^1{}_{ij}\S_3{}^j+R^2{}_{ij}\S_1{}^j)\bigr)
+\mu_i(\S_2{}^i-\a v^i)+\r(\a)^2.\cr} \eqno(2.14)
$$
In the above we have introduced $v^i:=\tv_{\2}{}^i$ and
$$\S_{\1}{}^i:=-\tv_{\1}{}^i,\quad\S_2{}^i:=-\a\tv_{\2}{}^i,\quad
\S_3{}^i:=-\tv_{\3}{}^i.           \eqno (2.15)
$$
The over (under) tilde indicates a density of weight plus (minus) one.  Thus,
$v^i \and \S_A{}^i$
are densities of weight one while the other variables are not densitized except
as indicated by the tilde.

It is evident from this form of the Lagrangian that only variation of $A^A{}_i
\and \S_A{}^i$ lead to dynamical equations.  The remaining variables lead to
equations on the hypersurface $t=$ constant, hence to constraint equations.
In the absence of coordinate or tetrad conditions, $N^i \and B^A$ are arbitrary
while $\a \and \mu_i$ are needed to make $t=$ constant a null surface and to
obtain the otherwise missing field equation.  The strange thing is that
${\ut N}v^i$ appear in that combination and cannot be solved for separately{}.
As long as the solution is complex, ${\ut N}$ is arbitrary.  But, when the
reality conditions are imposed, $v^i = {\bar \S}_3{}^i$ and {\ut N} will no
longer be undetermined.  This structure in the Lagrangian foreshadows the fact
that the scalar constraint is second class.

Therefore, we take the phase space to be defined by $(A^A{}_i,\ \S_A{}^i)$.  It
is easy to see that they form canonically conjugate pairs with the Poisson
brackets
$$
\{A^A{}_i(x),\ \S_B{}^i(x')\} \= \d^A{}_B \d^3(x-x').  \eqno (2.16)
$$
Then the Hamiltonian takes the form
$$
H \= \int d^3x\{{\ut N}\H_0 \pl N^i\H_i \- B^A\G_A \- \mu_iC^i \- \r (\a)^2\}
\eqno (2.17)
$$
where
$$  \eqalign{
\H_0&:=v^i\bigl(R^1{}_{ij}\S_3{}^j+R^2{}_{ij}\S_1{}^j\bigr)\=0,\cr
\H_i&:=-R^A{}_{ij}\S_A{}^j\=0,\cr
\G_A &:=D_i \S_A{}^i \=0, \cr
C^i&:=\S_2{}^i+\a v^i \=0.\cr} \eqno (2.18a)
$$
are constraints which arise from varying the action with respect to ${\ut N},
N^i, B^A, \and \break \mu_i$. respectively.  Other constraints come from
varying $\r, \a, \and v^i:$
$$
\a=0, \qquad v^i\mu_i=0, \eqno (2.18b)
$$
and
$$
\phi_i\ :=\  R^1{}_{ij}\S_3{}^j \pl R^2{}_{ij}\S_1{}^j\= 0. \eqno (2.18c)
$$
Propagation of the constraint $C^i=0$ leads to
$$
\chi^i:\=2{\d^{B}{}_2}D_j\Bigl(
{\ut N}v^{[i}\S_A{}^{j]}Q^A{}_B\Bigr)-2A^3{}_jN^{[i}\S_1{}^{j]}
-B^3\S_1{}^i\= 0,  \eqno(2.18d)
$$
$$
 Q^A{}_B\ :=\ \d^A_3 \d^1_B\ +\ \d^A_1 \d^2_B.
$$
These are all the constraints, but propagation of the constraint $\G_3=0$ leads
to a condition on $\mu_i$ which is important in obtaining the otherwise missing
equation:
$$
\mu_i\S_1{}^i = R^1{}_{i0}\S_3{}^i - R^1{}_{ij}\S_3{}^i N^j.  \eqno (2.19)
$$
Note that $v^i \phi_i \equiv \H_0 \and \S_1{}^i \phi_i \equiv \S_3{}^i \H_i$.

Thus, there are 14 constraints among the dynamical variables and three
conditions on the Lagrange multipliers $\a \and \mu_i$.  Five of the
constraints, $\H_i, \G_1, \and \G_2$ are first class while the remaining
constraints, including $\H_0$ are second class.  Three of the second class
constraints are conditions on $v^i$, so there are 16 conditions on the 18
phase space variables per hypersurface point.

The Hamiltonian equations of motion for the dynamical variables are
$$
A^1{}_{i,0}\= {\d^1{}_A}D_iB^A+N^jR^1{}_{ij}-{\ut N}v^jR^2{}_{ij},
\eqno(2.20a)
$$
$$
A^2{}_{i,0}\={\d^2{}_A}D_iB^A+N^jR^2{}_{ij}-\mu_i,\eqno(2.20b)
$$
$$
A^3{}_{i,0}\={\d^3{}_A}D_iB^A+N^jR^3{}_{ij}-
{\ut N}v^jR^1{}_{ij},\eqno(2.20c)
$$
$$
\S_1{}^i{}_{,0}\=2{\d^B{}_1}D_j\Bigl({\ut
N}v^{[i}\S_A{}^{j]}Q^A{}_B\Bigr)-2{\d^B{}_1}D_j(N^{[i}\S_B{}^{j]})-
2B^3\S_3{}^i,\eqno(2.20d)
$$
$$
\eqalign {\S_3{}^i{}_{,0}\=2&{\d^B{}_3}D_j\Bigl(
{\ut N}v^{[i}\S_A{}^{j]}Q^A{}_B\Bigr)-
2{\d^B{}_3}D_j(N^{[i}\S_B{}^{j]})\cr
&+2B^1\S_3{}^i+B^2\S_1{}^i,}\eqno(2.20e)
$$

The quantities we have introduced are complex.  To recover the Einstein theory
we must impose reality conditions at some point.  These conditions are that
$$
 {\bar \S}_1{}^i \= -\S_1{}^i, \qquad v^i \= {\bar \S}_3{}^i,
$$
and that
$$
 \Ga^1 -2\o^{23},\quad {\bar \Ga^2}=\o^{\3\1}, \quad\and {\bar \Ga^3}=\o^{\0\2}
$$
form the anti-self-dual components of the connection.
\vfill
\eject

\noindent 3.  Analysis of the Equations.

	We assume that space-time is asymptotically Minkowskian and that
 outside of a
timelike cylinder, the coordinate $t$ defines a congruence of hypersurfaces{}.
When $\a=0$, these are null surfaces which have the topology of null cones
extending to null infinity.  These null cones in turn are foliated by closed
two-surfaces so that each null generator is labeled by the usual angular
coordinates $(\theta, \phi)$ of the unit sphere.  Following Bondi and Sachs
[11,12], we choose the
coordinate $r$ along the generators to be the luminosity distance;  that is,
the area of each two surface $r=$ constant is $4\pi r^2$.  The coordinates
$x^i$ of the previous section are then $(r, \theta, \phi)$.  For the
boundary conditions, we assume that
$\S_A{}^i \and v^i$, thus the metric as well, take their Minkowski space
behavior in limit of null infinity.  In as far as it is possible to
do so without losing the
gravitational radiation, the same is true for the connection.  It turns
out that $A^{\2}{}_i$ contains radiative terms in this limit.  These
conditions, including the radiative terms, are unaffected by
super-translations.

The five first class constraints allow us to make a convenient choice for the
triad densities and the coordinates.  One coordinate condition has been used to
fix $r$ as the luminosity distance:
$$
  -ir^2\sin \theta\ \n_{1jk}\ v^j\S_{\3}{}^k  \= \nu^2. \eqno (3.1a)
$$
With the other two available coordinate conditions, we can set
$$
\S_{\1}{}^i \= -ir^2\sin \theta\  \d^i{}_1.  \eqno (3.1b)
$$
Then the four real functions in the null
rotations generated by $\G_1 \and \G_2$ allow us to fix $\S_{\3}{}^i$ tangent
to the surfaces $r=$constant, that is, $\S_{3}{}^1=0$, and then to set
$A^{\1}{}_1=0$.  The latter is equivalent to setting $\e=0$ in the
Newman-Penrose formalism [14].

As pointed at the end of the previous section, the Dirac brackets of the
conjugate pair $(A^{\3}{}_2,\ \S_{\3}{}^2)$ are equal to their Poisson
brackets.  Therefore, these can be identified as the dynamically independent
degrees of freedom for the gravitational field.  However, they are not
observables because they are not diffeomorphism invariant.  Nonetheless, they
 represent the initial
data which can be specified on an initial null surface.  The remaining
variables and parameters are then determined by any remaining gauge freedom,
the constraints, and relations from the propagation equations.

With the coordinate and gauge conditions given above, the constraints and
propagation equations have a natural order for their solution.  Below we will
give the equations in the order in which they can be solved.  Which variable
is to be solved for is generally clear.  (The indices $a,b,..$
range over 2 and 3.)
$$
\eqalignno{
\G_{\1}=0 \Rtar &  \S_{\1}{}^i{}_{,1}+
2A^{\3}{}_i\S_{\3}{}^i \= 0 &(3.2a) \cr
&\cr
\G_{\2}=0 \Rtar & A^{\3}{}_i\S_{\1}{}^i \= 0  & (3.2b) \cr
&\cr
\H_1=0 \Rtar & A^{\3}{}_{i,1} \S_{\3}{}^i \= 0
&  (3.2c) \cr
& \cr
\G_{\3}=0 \Rtar & \S_{\3}{}^i{}_{,i}- 2A^{\1}{}_i\S_{\3}{}^i
- A^{\2}{}_i\S_{\1}{}^i \= 0 &  (3.2d1) \cr
& \cr
\H_a=0 \Rtar &  [A^{\1}{}_{a,j}- A^{\1}{}_{j,a} -
A^{\2}{}_a A^{\3}{}_j + A^{\3}{}_a A^{\2}{}_j]\S_{\1}{}^j \cr
&+ [A^{\3}{}_{a,j}- A^{\3}{}_{j,a} +
A^{\3}{}_a A^{\1}{}_j - A^{\3}{}_j A^{\1}{}_a]\S_{\3}{}^j \= 0  \cr
&[A^{\1}{}_{a,j}- A^{\1}{}_{j,a} -
A^{\2}{}_a A^{\3}{}_j ]\S_{\1}{}^j +
A^{\3}{}_a [\S_{\3}{}^i{}_{,i}- 2A^{\1}{}_i\S_{\3}{}^i] + \cr
& [A^{\3}{}_{a,j}- A^{\3}{}_{j,a} +
A^{\3}{}_a A^{\1}{}_j - A^{\3}{}_j A^{\1}{}_a]\S_{\3}{}^j \= 0
&  (3.2d2) \cr
& \cr
\phi_a=0 \Rtar & A^{\2}{}_{a,j}\S_{\1}{}^j -
\S_{\3}{}^c A^{\3}{}_c A^{\2}{}_b Z^b{}_a + G_a \= 0  & (3.2e) \cr
& Z^b{}_a:= \d^b{}_a-{{\S_{\3}{}^b A^{\3}{}_a}\over{\S_{\3}{}^c A^{\3}{}_c}},\cr
& G_a:=(A^{\1}{}_{a,b}-A^{\1}{}_{b,a})\S_{\3}{}^b -
(A^{\2}{}_{j,a}+2A^{\2}{}_jA^{\1}{}_a)\S_{\1}{}^j \cr
& \cr
\chi^a=0 \Rtar & (V^a \S_{\1}{}^j)_{,j} +
\S_{\3}{}^c A^{\3}{}_c Z^a{}_b V^b\= 0, &  (3.2f) \cr
& {\ut N}v^a\equiv V^a \cr
& \cr
\chi^1=0 \Rtar & B^{\3} \S_{\1}{}^1 + (V^b\S_{\1}{}^1)_{,b} +
2 A^{\1}{}_b V^b \S_{\1}{}^1 - A^{\3}{}_b N^b \S_{\1}{}^1 \= 0
&  (3.2g1) \cr
& \cr
{\dot \S}_{\1}{}^a=0 \Rtar & (N^a \S_{\1}{}^j)_{,j} -
(V^a\S_{\3}{}^b-V^b\S_{\3}{}^a)_{,b}+ 2A^{\2}{}_j\S_{\1}{}^j  \cr
&+  2A^{\3}{}_b\S_{\3}{}^b N^a + 2 B^{\3} \S_{\3}{}^a =  0 &  (3.2g2)\cr
& \cr
{\dot \S}_{\1}{}^1=0 \Rtar & 2N^1 A^{\3}{}_b\S_{\3}{}^b -
A^{\2}{}_jV^j\S_{\1}{}^1 - (N^a\S_{\1}{}^1)_{,a} \= 0  &  (3.2h) \cr
& \cr
{\dot A}^{\1}{}_1=0 \Rtar & B^{\1}{}_{,1} + A^{\2}{}_1 B^{\3} -
N^a(A^{\1}{}_{a,1} + A^{\3}{}_a  A^{\2}{}_1) \cr
& - V^a(A^{2}{}_{1,a- A^{\2}{}_{a,1}}- 2A^{\1}{}_a A^{\2}{}_1) \= 0
&  (3.2i) \cr
& \cr
{\dot \S}_{\3}{}^1 =0 \Rtar & B^{\2} \S_{\1}{}^1 -
(N^1\S_{\3}{}^a)_{,a} + 2N^1A^{\1}{}_a\S_{\3}{}^a \cr
& -N^a A^{\2}{}_a \S_{\1}{}^1 \= 0 &  (3.2j) \cr
}
$$
This completes the set of equations which can be solved either by
integration along the null generators of the surface $t=$constant or
algebraically on that surface.
The equation for${\dot A}^{\3}{}_1=0$ yields another equation for $B^{\3}$
which is then trivially satisfied.

The remaining equations all contain time derivatives.
$$
\eqalignno{
{\dot A}^{\1}{}_a \= &B^{\1}{}_{,a}+ A^{\2}{}_a B^{\3}- A^{\3}{}_a B^{\2} +
N^j[A^{\1}{}_{a,j}- A^{\1}{}_{j,a}+ A^{\2}{}_j A^{\3}{}_a -
A^{\3}{}_j A^{\2}{}_a]  \cr
&-V^b[A^{\2}{}_{a,b}- A^{\2}{}_{b,a}+ A^{\1}{}_b A^{\2}{}_a -
A^{\2}{}_b A^{\1}{}_a] & (3.3a) \cr
& \cr
{\dot A}^{\2}{}_i\= &B^{\2}{}_{,i} - 2A^{\1}{}_iB^{\2}+ 2A^{\2}{}_iB^{\1} \cr
&+ N^j[A^{\2}{}_{i,j}- A^{\2}{}_{j,i}- 2A^{\1}{}_j A^{\2}{}_i +
2A^{\2}{}_j A^{\1}{}_i] - \mu_i & (3.3b) \cr
& \cr
{\dot A}^{\3}{}_a \=& B^{\3}{}_{,a} - 2A^{\3}{}_a B^{\1} +
2A^{\1}{}_a B^{\3} + N^j [A^{\3}{}_{a,j}- A^{\3}{}_{j,a}+
2A^{\1}{}_j A^{\3}{}_a - 2A^{\3}{}_j A^{\1}{}_a] \cr
&- V^b[A^{\1}{}_{b,a}- A^{\1}{}_{a,b}+ A^{\2}{}_a A^{\3}{}_b -
A^{\3}{}_a A^{\2}{}_b] & (3.3c) \cr
& \cr
{\dot \S}_{\3}{}^a \=& -2\rd_j(N^{[a}\S_{\3}{}^{j]}) +
4A^{\1}{}_jN^{[a}\S_{\3}{}^{j]} + 2A^{\2}{}_jN^{[a}\S_{\1}{}^{j]} \cr
&-2B^{\1} \S_{\3}{}^a + B^{\2}\S_{\1}{}^a & (3.3d) \cr
}
$$

The fact that the Poisson brackets of the constraints with the Hamiltonian
vanish modulo the constraints themselves tells us that the propagation of the
variables is determined by the propagation of $(A^{\3}{}_2,\ \S_{\3}{}^2)$
alone.  That is not quite true because, as in the case of Bondi-Sachs [11,12]
 and
Newman-Penrose [13], integration along the null generators introduces arbitrary
functions of $(t,\theta,\phi)$.  We find that propagation of $A^{\1}{}_a \and
A^{\2}{}_a v^a$  lead to the conservation equations for angular momentum and
mass.  The time derivative of $A^{\2}{}_a\S_{\1}{}^a $ defines
$\mu_i\S_{\1}{}^i$ which when equated to (2.19) yields an identically
satisfied field equation.  On the other hand, the time derivative of
$A^{\2}{}_a\S_{\3}{}^a $ defines $\mu_i\S_{\3}{}^i $ as the null component of
the conformal tensor.  This is in complete agreement with the previous work{}.

The integration along the null generators can be carried out from the time-like
cylinder to null infinity without an expansion in $1/r$ in a manner
similar to that of Tamburino and Winicour [21], but this
formal result does not exhibit the presence or lack of
logarithmic behavior.  Therefore, in
the next section we shall set up the calculation of the asymptotic behavior and
then list the results.
\vfill
\eject

\noindent 4.  The Asymptotic Solution. \hfill\break

In this section we shall first solve the complex equations for the triad and
the self-dual connection.  Then we shall apply the reality conditions and show
the explicit relationship of our results to those of Sachs [12].
We shall look for solutions for which the triad differs from its Minkowski
space value by factors with an
expansion in $1/r$.  Although there are logarithmic terms consistent with
the assumption of
asymptotically Minkowskian behavior, in this section we shall choose the
powers of  $1/r$ to avoid their
occurrence.  The appearance of terms in $(\ln\,r)^m/r^n$ will be discussed
in the following section.  The Minkowski space solutions are given in
Appendix 1.

\noindent {\it The Solution.}

As noted previously, the initial data on the surface $t=0$ will be given by
$(A^{\3}{}_2,\ \S_{\3}{}^2).$
By our coordinate conditions we have
$$
 \S_{\1}{}^i \= -ir^2 \sin \theta\  \d^i{}_1. \eqno (4.1)
$$
Then, $\G_{\1} \and   \G_{\2}$, Eqs.(3.2a,b) tell us that
$$
\eqalignno{
\S_{\3}{}^a A^{\3}{}_a &\=  ir\sin \theta, & (4.2a) \cr
A^{\3}{}_1&\=0.  & (4.2b) \cr
}
$$
To proceed, we write
$$
\eqalign{
A^{\3}{}_2 &\= -{\Io{\rt 2}}\{1+ {{{}^3\!A}\over {r^3}}+\cdots .\} \cr
\S_{\3}{}^2  &\= -{{ir\sin \theta }\over \rt2}\{ 1 +  {{{}^1\!\S}\over r} +
{{{}^3\!\S}\over {r^3}}+\cdots .\} \cr} \eqno (4.3)
$$
and find from  (4.2a) and $\H_1$ that
$$
\eqalign{
A^{\3}{}_a  &\= -a_a + - { {{}^3\!A}\over{r^3}} b_a  \cr
\S_{\3}{}^a  &\= -ir\sin \theta \{ s^a + { {{}^1\!\S}\over r}t^a +
{{{}^3\!\S}\over {r^3}}t^a \}.\cr }
\eqno (4.4)
$$
The forms $(a_a, b_a) $ and the vectors $(s^a, t^a)$ are the eigen-forms and
eigenvectors of the flat space
part of the matrix $Z^a{}_b$ as defined in Appendix 2.  They satisfy
the following algebraic relations:
$$
a_a s^a \= b_a t^a \= 1, \qquad a_a t^a \= b_a s^a \= 0.
$$
We  assign a spin-weight of -1 to $b_a \and s^a$ and a spin-weight of +1 to
$a_a \and t^a$ [22-24] .  This
allows us to express the results in terms of spin-weighted quantities which act
 as a check on the
calculations and simplifies their appearance through the use of the edth
operator which is also defined in
the Appendix.  In (4.4) and below, we exhibit only the terms of the solution
we have calculated and omit
the dots indicating further terms.

The radial integrations are not unique, but lead to the introduction of a
number of $r-$independent functions, $C^{\1}{}_s=C^{\1}{}_a s^a,
C^{\1}{}_t=C^{\1}{}_a t^a, \M, \A,
{}^0\!\V, \and {}^1\!\V.$  The first four of these functions are related to the
angular momentum, mass, and radiation.  The remaining two are fixed by the
reality conditions.
Below we give the results of these radial integrations in the order
given in Eqs. (3.2):
$$
\eqalignno{
A^{\1}{}_a  \= & \bigl[{{\cot \theta}\over {2\rt2}} - {\Io r}\edth\,
{{{}^1\!\S} } + {\Io r^2}C^{\1}{}_s -
{\Io r^3}{\edth\, {{}^3\!\S} } \bigr] a_a + \cr
&\bigl[-{{\cot \theta}\over {2\rt2}} + {\Io r^2}C^{\1}{}_t -
{\Io r^3} {\bedth\,} {{}^3\!A} \bigr] b_a
& (4.5a) \cr
& \cr
A^{\2}{}_1 \= &{\Io r^2}\bigl[ {\edth\, {{}^1\!\S} } - {\Io r}C^{\1}{}_s  +
{\Io r^2} ({\edth\, {{}^3\!\S} } - C^{\1}{}_t)\bigr] & (4.5b) \cr
& \cr
A^{\2}{}_a  \=  &\bigl[ \A + {\Io r}\bedth\,\edth\, {{}^1\!\S}  -
{\Io {2r^2}}(\M\,{{}^1\!\S}  +
\bedth\, C^{\1}{}_s + 2(\edth\, {{}^1\!\S})^2 + {{}^1\!\S} \edth^2\,{{}^1\!\S}
\bigr] a_a \cr
& + \bigl[-{\Io 2} + {\Io r}\M - {\Io r^2} \bedth\, C^{\1}{}_t \bigr] b_a
&  (4.5c) \cr
& \cr
V^a \= &-{\Io r}\Bigl[({\Io r}{}^1\!\V + {\Io {2r^3}} {}^0\!\V ) s^a  +
 ( {}^0\!\V{{}^3\!A} + {\Io {2r^2}} {}^1\!\V {{}^1\!\S} ) t^a \Bigr]
& (4.5d) \cr
& \cr
B^{\3} \=  & O({\Io {r^3}})  & (4.5e) \cr
& \cr
N^a \= & -{\Io r^2}\bigl[ \bedth\, {}^1\!\V s^a  +
{}^0\!\V \edth\, {{}^1\!\S} t^a
\bigr] & (4.5f) \cr
& \cr
N^1 \= & -{\Io 2}\bigl\{ {}^0\!\V  - {\Io r}[ 2{}^0\!\V \M +
\bedth^2\,{}^1\!\V  +
\edth\, ( {}^0\!\V \edth\, {{}^1\!\S} +2\A{{}^1\!\V})] \bigr\} &(4.5g) \cr
& \cr
B^{\1} \= & {\Io {2r^2}} {}^0\!\V (\edth^2 \,{{}^1\!\S} - \M) &(4.5h) \cr
& \cr
B^{\2} \= & {\Io r^2}\bedth \bigl\{ ({}^0\!\V \M) +
{\Io 2}[\bedth^2 \,{}^1\!\V  +
\edth ( {}^0\!\V \edth\, {{}^1\!\S} ) ]  - {}^1\!\V \bedth  \A \bigr\}
&(4.5i) \cr
}
$$
The integration for $N^a$ can introduce a function independent of $r$.
However, such a term can be removed by a coordinate transformation [11,12].
Furthermore, $N^1$ will grow like $r$ at null infinity unless
we require that
$$
\edth\, {}^0\!\V \= \bedth\, {}^0\!\V \= 0.
$$
The above solutions have been written with this requirement so
that at most, $ {}^0\!\V $ can be a function of $t$ alone.

The fact that the constraints form a closed system shows that the propagation
of these equations is
consistent.  This means that the propagation equations will determine the
evolution of the arbitrary
functions we have introduced, but there will be no further conditions.  This
argument is equivalent to the
use of the Bianchi identities by Bondi and Sachs [11, 12] to obtain a
similar result.
$ {{}^3\!A} , {{}^1\!\S} ,
\and {{}^3\!\S} $ are part of our initial data.  In addition we have
$C^{\1}{}_a , \M, \A,  {}^0\!\V, \and
{}^1\!\V $.   These are exactly the same quantities we would have had to
introduce if we were to integrate
the equations without the asymptotic expansion.

{}From the evolution equation (3.3a), we obtain the following relations:
$$
{}^0\!\V \A  \= {\dot  {{}^1\!\S} } \eqno (4.6a)
$$
$$
\eqalignno{
{\dot C}_s \=  & -\bedth\, {}^2\!B^{\1} + {}^2\!B^{\2} +
{}^0\!\V \bigl[ -\edth ( \bedth\edth\, {{}^1\!\S}  - {\Io 2} {{}^1\!\S} )
+ \bedth \M ]   & (4.6b) \cr
& \cr
{\dot C}_t \= & -\edth\, {}^2\!B^{\1}  + {}^1\!\V \bedth \A   & (4.6c) \cr}
$$
These latter equations are identified with the change in dipole aspect and,
hence, are connected with the
conservation of angular momentum.   The equation for
${\dot A}^{\2}{}_i \S_{\1}{}^i $ defines $\mu_1$
which leads to an identity with Eq. (2.18).  $\mu_i \S_{\3}{}^i $ is the
null part of the conformal tensor,
$\Psi_4$ in the Newman-Penrose notation [13].  Thus, only the equation
for ${\dot A}^{\2}{}_a V^a$ is
dynamical.  Remembering that $\mu_i v^i=0$, the relevant relation
can be written as
$$
-{\rd\over {\rd t}}\bigl( {}^0\!\V \M - {}^1\!\V {}^1\!{\dot \S}\bigr) \- {}^1\!{\dot \V} {}^1\!{\dot \S}
\eqno (4.7)
$$
All that remains now are the equations to propagate $ A^{\3}{}_a
\and \S_{\3}{}^a $.  This completes
the solution of the complex equations.  The further discussion of (4.7)
will be delayed until after the reality conditions have been applied.

\noindent {\it The Reality Conditions.}

It is only necessary to apply the reality conditions to the tetrad because
through the solution of the field equations the connection is expressed
in terms of the tetrad.  We shall see that in applying the reality conditions,
the arbitrary functions in the connection will be expressed in terms of those
in the triad.  The reality conditions on the tetrad are
$$
N\= {\bar N}, \qquad N^i\= {\bar N}^i, \qquad -i\nu\= i{\bar \nu}\ >0,
$$
$$
\Sa^i\= -{\bar \Sa^i}, \qquad v^i\= {\bar \Sc^i}. \eqno (4.8)
$$

{}From (3.1) we find that $\Sa^1$ already satisfies the reality condition.
Writing $v^a={\bar \Sc}$, we have
$$
\eqalign{
\Sc\= &-ir\sin\theta\{s^a+{\Io r}{}^1\!\S t^a+ {\Io r^3}{}^3\!\S t^a\},\cr
v^a\= &\ ir\sin\theta\{t^a+{\Io r}{}^1\!{\bar \S}s^a+
{\Io r^3}{}^3\!{\bar \S}s^a\}. \cr
}\eqno (4.9)
$$
{}From this we find that
$$
\nu^2 \= \n_{ijk}\Sa^i v^j \Sc^k\-r^4\sin^2\theta\bigl[1-{\Io r^2}{}^1\!\S{}^1\!{\bar \S}\bigr].
$$
The requirement $-i\nu >0$ gives us
$$
\nu\= ir^2\sin\theta\sqrt{1-{\Io r^2}{}^1\!\S{}^1\!{\bar \S}}. \eqno (4.10)
$$
Now from ${\ut N}v^a=V^a$ and the boundary condition that $N$ should be 1 at
null infinity, we find
$$
\eqalign{
{N\over r}&\bigl(1+{\Io 2r^2}{}^1\!\S{}^1\!{\bar \S}\bigr)
\bigl[t^a+{\Io r}{}^1\!{\bar \S}s^a+ {\Io r^3}{}^3\!{\bar \S}s^a\bigr] \cr
&=\ -{\Io r}\bigl[\bigl({}^0\!\V+{\Io 2r^2}{}^1\!\V{}^1\!\S\bigr)t^a\pl
\bigl({\Io r}{}^1\!\V\pl {\Io 2r^3}{}^0\!\V\ {}^3\!A\bigr)s^a,
}
$$
which implies $N=1$ and
$$
{}^0\!\V=-1,\qquad {}^1\!\V=-{}^1\!{\bar \S},\qquad
{\Io 2}{}^3A= {}^3\!{\bar \S}+{\Io 2}{}^1\!\S{}^1\!{\bar \S}{}^1\!{\bar \S}.
\eqno (4.11)
$$

Comparison with the Bondi-Sachs form of the metric [12,22]
$$
ds^2\= {Ve^{2b}\over r}du^2+2e^{2b}dudr-
r^2h_{ij}(dx^i-U^idu)(dx^j-U^jdu),
$$
give us
$$
\nu\= -ie^{2b}r^2\sin\theta,\qquad N\=1, \qquad N^1\= {V\over r},
$$
and for the principal spin coefficients
$$
\eqalign{
\r\= &-{\Io \nu} \Ac_a\Sc^a\= -{1\over r}-{{}^1\!\S{}^1\!{\bar \S}\over r^3},\cr
\s\= &{\Io \nu}\Ac_av^a\= -{{}^1\!{\bar \S}\over r^2}+{\Io r^4}({}^3\!A-
{\Io 2}{}^1\!\S{}^1\!{\bar \S}{}^1\!{\bar \S}-{}^3\!{\bar \S}),\cr
\mu\= &{\Io \nu}\Ab_av^a\= -{\Io 2r}+
{\Io r^2}(\M -{}^1\!{\dot \S}{}^1\!{\bar \S}), \cr
{\bar \l}\= &-{\Io \nu}\Ab_a\Sc^a\= -{{}^1\!{\dot \S}\over r}+
{\Io r^2}(\bedth\edth {}^1\!\S-{\Io 2}{}^1\!\S).\cr
}\eqno (4.12)
$$
This shows that the Newman-Penrose asymptotic shear $\s^0=- {}^1\!{\bar \S}$
and
$$
-\M+{}^1\!{\dot \S}{}^1\!{\bar \S}\= {\Io 2}\bigl[\psi_2^0+{\bar \psi}_2^0+
\bedth^2 \s^0+\edth^2 {\bar \s}^0+
{}^1\!{\dot \S}{}^1\!{\bar \S}+ {}^1\!{\dot {\bar \S}}{}^1\!\S\bigr].
\eqno (4.13)
$$

Note that the right hand side of (4.13) is real and comparison with (4.7) shows
that it is just the negative of the Bondi-Sachs mass aspect.

It is perhaps wothwhile to exhibit the real mass loss expicitly by writing
the integral over a sphere at null infinity as
($dS$ is the area element of the unit sphere)
$$
{{dM}\over {dt}} \-{\Io {16\pi}}\oint {}^1\!{\dot \S}\,{}^1\!{\dot {\bar \S}} dS, \eqno (4.14)
$$
$$
\eqalign{
M \ :=\ &{\Io {16\pi}}\oint \bigl[\M + {\bar \M} -
{}^1\!{\bar \S}{}^1\!{\dot \S} -
{}^1\! \S {}^1\!{\dot {\bar \S}}\bigr] dS \cr
0\=&\oint \bigl[\M - {\bar \M} - {}^1\!{\bar \S}{}^1\!{\dot \S} +
{}^1\! \S {}^1\!{\dot {\bar \S}}\bigr]  \cr } \eqno (4.15)
$$
This definition of the mass aspect agrees with that of Bondi-Sachs and
Newman-Penrose.  Thus, (4.14)
describes the mass loss from gravitational radiation.
It is interesting to note that addition of the surface integral
$$
\oint V^a A^{\2}{}_a  \S_{\1}{}^1 d\theta d\phi \eqno (4.16)
$$
is needed to assure the differentiability of the Hamiltonian.  Together with the
reality conditions, (4.16) defines the mass $M$.
\medskip

\noindent  5.  Logarithmic Behavior.  \hfill \break

In this section we shall consider the terms in $1/r$ which were omitted in
the definition of
$( A^{\3}{}_a , \S_{\3}{}^a )$ in the previous section. These terms lead
to logarithmic behavior which
comes in below the leading terms previously found.   Chrusciel, MacCallum, and
Singleton  [15] have studied
the logarithmic behavior within the Bondi-Sachs formalism [11, 12].  They
introduce polyhomogeneous
functions in the metric and then see what powers of $(\ln r)^m/ r^n$
lead to consistency in the solution of
the constraint and propagation equations.  Here we follow a somewhat
different approach.  We start with a
power series in $1/r$ in $( A^{\3}{}_2 , \S_{\3}{}^2 )$ and examine the
logarithmic terms which arise in
the remaining terms.  We then look at the propagation equations to see
what logarithmic terms are
introduced into
$( A^{\3}{}_a , \S_{\3}{}^a )$.  Consistency means that these new
terms should not interfere with what
has been previously required.  That is, a new calculation with these
new terms should reproduce the
already found behavior and add lower order logarithmic behavior. Apart
from this difference in approach,
our results are essentially in agreement with those of Chrusciel, MacCallum, and
Singleton, but perhaps they are more perspicuous.  Our approach is closer in
spririt to that of Winicour [25].  However, in our results the coefficient
of $1/r$ in $\Ac_a$ is found to be zero, whereas that does not appear to be the
case in their analysis.  For possible clarification, the details of the
calculation leading to that result is given in Appendix 3.
Below we sketch our final results and then discuss the conservation
equations and the conformal tensor.

{}From $\G_{\1}, \G_{\2}, \and \H_1$ we find
$$
\eqalign{
A^{\3}{}_a  &\= -a_a \pl {{{}^2\!A}\over  r^2}  b_a  \pl
{{{}^3\!A}\over  r^3}   \cr
\S_{\3}{}^a  &\= -ir\sin \theta \Bigl[ s^a \pl
{{{}^1\!\S}\over r}t^a \pl { {{}^2\!\S}\over r^2}t^a \pl
{{{({}^3\!\S}-{{}^1\!\S}{{}^2\!A})} \over {r^3}} \Bigr].\cr
}   \eqno (5.1)
$$
The later integration for $V^a$ requires that a possible
term in $1/r$ in $ A^{\3}{}_a$
 be set equal to zero.   Then the remaining integrations along the
null generators give
$$
\eqalignno{
A^{\1}{}_a  \= & \Bigl[{{\cot \theta}\over {2\rt2}} \- {\Io r}\edth\,
{{{}^1\!\S}}
\pl \edth\, {{}^2\!\S}{{\ln r}\over  r^2} \pl {\Io r^2}C^{\1}{}_s \Bigr]
a_a \pl  \cr
&\Bigl[-{{\cot \theta}\over {2\rt2}} + \bedth\, {{}^2\!A} {{\ln r}\over  r^2}
\pl {\Io r^2}C^{\1}{}_t \ \Bigr] b_a     & (5.2a)  \cr
A^{\2}{}_1 \= &{\Io r^2}\Bigl[ {\edth\, {{}^1\!\S} } \pl
2\edth\, {{}^2\!\S}\ {{\ln r}\over r}\Bigr]
& (5.2b) \cr
& \cr
A^{\2}{}_a  \=  &\Bigl[- {}^1 \!{\dot\S} \pl
{\Io r}\bedth\edth\,{{}^1\!\S}  \pl
{{}^1\!\S} \bedth\edth\,{{}^2\!\S}\ {{\ln r}\over  r^2} \pl
{\Io r^2} {{}^2\!A}  \Bigr] a_a \cr
& \pl \Bigl[-{\Io 2} \pl {\Io r}\M -
{{\ln r}\over r^2}\bigl(\bedth^2 \,{{}^2\!A} -\edth^2 \,{{}^2\!\S}\bigr)
\Bigr]
 b_a &  (5.2c) \cr }
$$
To  the order considered,  the solution for $V^a $  is the same as in the
previous section.  However, as
noted above, it imposes the condition that $ {{}^1\!A} =0$.
This result is contained in the above expressions.

In the remaining variables, the logarithms appear in the order below the
leading order given in the
previous section. Except for $N^1$, they have no important
consequences.  For $N^1$, we find  a term in
${{\ln r}/r^2}$.  It then follows from the propagation
equations that $ A^{\3}{}_a  $ develops a term in
${\ln r}/r^3 \and \S_{\3}{}^a $ a term in
${\ln r}/r^2$ which follows the term in $ {{}^1\!\S} $ .
As a result, the conservation equation for mass is
unchanged.  This means that what appears as gravitational
radiation at null infinity is unaffected by the
inclusion of this logarithmic behavior.  On the other hand, because
$ A^{\1}{}_a $ has a term in
$ {\ln r}/r^2$, the conservation equation for angular momentum will be
changed.  This represents another
problem for angular momentum which is yet to be
understood adequately.  This is exhibited below by the
failure of the conformal tensor to peel:
$$
\eqalign{
\Psi_4\= & R^{\2}{}_{\r\s}e_{\0}{}^{\r} e_{\3}{}^{\s} \rtar
{{}^1\!\ddot\S\over r }, \cr
\Psi_3\= & R^{\1}{}_{\r\s}e_{\3}{}^{\r} e_{\0}{}^{\s} \rtar
{-\edth\, {}^1\!\dot\S\over r^2}, \cr
\Psi_2\= & R^{\2}{}_{\r\s}e_{\1}{}^{\r} e_{\2}{}^{\s} \rtar
{{\edth^2 \,{{}^1\!\S} - \M}\over r^3}, \cr
\Psi_1\= & R^{\1}{}_{\r\s}e_{\1}{}^{\r} e_{\2}{}^{\s} \rtar
 2\bedth\ {{}^2\!A} {{\ln r}\over r^4}
\pl {{ 2C^{\1}{}_t \- \bedth\ {{}^2\!A}}\over r^4}, \cr
\Psi_0 \= & R^{\1}{}_{\r\s}e_{\1}{}^{\r} e_{\2}{}^{\s} \rtar
 -{ {{}^2\!A}\over r^4}
- 3\a{{\ln r}\over r^5} \- {{ {{}^3\!A} \- \a}\over r^5}.\cr
}  \eqno (5.3)
$$
In the above, the $  \Psi_n$ are the components of the conformal tensor,
 ${{}^1\!\S} = -{\bar\s}^{\circ}$ of Newman-Penrose, and  $\a$
comes from the lowest order
logarithm in $ A^{\3}{}_a $.

The difference between our results and those of Chrusciel, MacCallum, and
Singleton [15] comes from the
different question which is asked.  We ask for the logarithmic behavior
which is forced on us by adding
the additional terms in $1/r$ in both $ A^{\3}{}_a \and \S_{\3}{}^a $
 while they ask for the most general
logarithmic behavior they can introduce into the metric which is
self-consistent.  As a result, they find
that the metric can have terms in ${(\ln r)^N\over r}$ whose coefficients
are independent of $u$.
However, the main physical conclusions are the same.  The definition of
Bondi mass and the radiation of
gravitational energy remains unchanged from the results without logarithmic
behavior.  Furthermore,
considerations about angular momentum are affected by these new terms.  On
the other hand, the current
results are in complete agreement with those of Novak and Goldberg [16, 17] who
showed that the existence of
null infinity was consistent with the logarithmic behavior found here.
\vfill
\eject

\noindent 6.  Conclusions.

One of the main points of this paper has been to see whether the new
variables introduced by
Abhay Ashtekar [1-4] are useful in classical problems.  Indeed,  use of
the self-dual connection and
the densitized triad $(A^{\A}{}_a , \S_{\A}{}^a)$ in the canonical
formallism leads to a set of
equations which are intermediate between the Bondi-Sachs [11, 12] and
Newman-Penrose [13] equations
in the vicinity of future null infinity,
${\cal I}^+$.  In Bondi-Sachs, the calculation begins with specification
of the metric on a two-surface foliation of an outgoing
asymptotic null cone;  in Newman-Penrose, with the specification
of a component of the conformal tensor, $\Psi_0{\ut N}^2 C_{ijkm} \S_{\1}{}^i \S_{\3}{}^j
\S_{\1}{}^k \S_{\3}{}^m$;  and in the present calculation we give as
our initial data
$( A^{\3}{}_2, \S_{\3}{}^2)$ which is only half of the two-surface
metric and that part of the
connection which is related to ${\bar\Psi}_0$.  We have not evaluated
whether our calculation is
the most efficient.  Both Newman-Penrose and we work with first order
equations.  Eventually
they make use of the ``metric'' equations to determine the tetrad
whereas they are part of our
canonical equations, but we must then compute the conformal tensor.

The one advantage of the present approach is that it is derived from a
Lagrangian and a
canonical formalism which may yet be useful for quantum gravity.  It also
makes the study of the
logarithmic behavior in the vicinity of null infinity somewhat more imperative
and somewhat
easier.  Apart from the point mentioned earlier and elaborated on in Appendix
3, the important conclusions we have arrived at are not significantly
different from those of
Chrusciel et al [15], if less complete.  We hope that this difference can be
resolved in the near future.  As noted in the previous section, we
ask different questions.
The important results here and there are that one can have an asymptotically
Minkowskian
metric, with a future null infinity,  and a mass and radiation of gravitational
 energy which is well
defined.  That is, the logarithmic terms fall off faster than those terms which
 define the mass and
radiation of gravitational energy.  The same is not true for angular momentum.
 But, that concept
is not sufficiently clear even in the absence of the logarithmic terms,
although, in that case, there
is an expression for angular momentum which transforms properly under the
super translations
as well as the Lorentz transformations [25].

There is, in the present approach, the additional need to apply the reality
conditions.  However,
note that there are no propagation equations for $v^i$.  Therefore, once
$v^i$ is identified with
${\bar \S}_{\3}{}^i $ , it follows that its propagation is also specified.
Given that
$ \S_{\1}{}^i $ is real and independent of time, the metric will propagate
as real.  That
guarantees that all the reality conditions will be fulfilled.

Our original hope was that on the null surface we could carry out a reduced
phase space
quantization of general relativity.  While the identification of the
dynamical degrees of freedom
for the gravitational field is easy, either one has to express all the
remaining variables in terms of
these degrees of freedom or construct the Dirac brackets.   At this time,
it appears to be very
difficult to carry out that task.

\noindent  Acknowledgement.

The authors wish to express their appreciation to David Robinson for his
readings of the several
drafts of this work.  They also thank Piotr Chrusciel for a critical reading of
the paper and an extended exchange in an attempt to understand the one apparent
difference in our results.
CS thanks the Department of Physics and the
Relativity Group of  Syracuse
University for its hospitality during the period when this work was begun.
JNG thanks the
faculty and staff of the Department of Mathematics, Kings College London for
 their hospitality
during June to December 1993 while part of this work was carried out.  He
also thanks John
Madore for discussions and the physics department for its hospitality during
a brief stay at the
Universit\'e de Paris-sud.
This work was supported in part by the NSF under Grant No. PHY 9005790 and
by SERC under Grant No. GR/H456472.
\vfill
\eject

\noindent Appendix 1.  Minkowski Space Tetrad and Connection. \hfill \break

In Minkowski space the metric takes the form
$$
ds^2 \= dt(dt \pl 2dr) \- r^2 (d\theta^2 \pl \sin^2 \theta\ d\phi^2)
$$
so that the four one-forms and tetrad vectors are
$$
\vcenter{
\halign{$\hfil # $ & ${}#\hfil $ &\qquad $\hfil #$ & ${}#\hfil $ \cr
\theta^{\0} \= & dt ,   & e_{\0} \= &\rd_t  \- \h\rd_r ,    \cr
\theta^{\1} \= &  (dr \pl \h dt),  & e_{\1} \= & \rd_r  \cr
\theta^{\2} \= & {\Io \rt2}r(d\theta \- i\sin \theta\ d\phi),
& e_{\2} \= & {\Io \rt2} (\rd_{\theta}
\pl {i\over {\sin\ \theta}}\rd_{\phi}) \cr
\theta^{\3} \= & {\Io \rt2}r(d\theta \pl i\sin \theta\ d\phi),
& e_{\3} \= & {\Io \rt2} (\rd_{\theta}\-
{i\over {\sin\ \theta}}\rd_{\phi}) \cr}}  \eqno (A1.1)
$$

Therefore, the triad densities and the self-dual connection one-forms are
$$
\vcenter{
\halign{$\hfil # $ & ${}#\hfil $ &\qquad $\hfil #$ & ${}#\hfil $ \cr
\S_{\1}{}^i &= -ir^2\sin \theta\ \d^i{}_1; & A^{\1}{}_i &{i\over 2}\cos\theta \d^i{}_3;  \cr
v^a & = ir \sin \theta \ t^a; & A^{\2}{}_a & = -{\Io 2}b_a ; \cr
\S_{\3}{}^a &= -ir \sin \theta\  s^a;  & A^{\3}{}_a & = -a_a; \cr}}
\eqno (A1.2)
$$
$$
s^a\= {\Io \rt2}\pmatrix{1\cr {-i/{\sin \theta}}\cr};\qquad
t^a\=  {\Io \rt2}\pmatrix{1\cr {i/{\sin \theta}}\cr};
\eqno (A1.3a)
$$
$$
\eqalign{
a_a\=&{\Io \rt2}(1,\ i\sin \theta);\cr
b_a\= &{ \Io \rt2}(1,\ -i\sin \theta).  \cr
}\eqno (A1.3b)
$$

The duality relations between $s^a , t^a \and a_a, b_a$,
respectively, are unchanged by a phase change
$e^{-i\phi}s^a, e^{i\phi}t^a \and e^{i\phi}a_a, e^{-i\phi}b_a$.
We assign a spin-weight of $-s$ to those
quatities which transform as $ e^{-is\phi}$ and a spin-weight of
$+s$ to those which transform as
$ e^{is\phi}$ under this possible change of  phase.  Keeping track
of the spin-weight is a help in
controlling the calculations  [22-24].

This also leads to the introduction of spin-weight raising and lowering
operators $\edth \and \bedth$.  If
$\n$ has spin-weight $s$, then
$$
\edth \n := \ -{\Io \rt2}\sin^s \theta \Bigl({\rd\over {\rd \theta}}\pl
{i\over {\sin \theta}}
{\rd\over {\rd \phi}}\Bigr)(\sin^{-s} \theta\,\n)  \eqno (A1.4a)
$$
and
$$
\bedth \n := \ -{\Io \rt2}\sin^{-s} \theta\Bigl({\rd\over {\rd \theta}}
\pl {i\over {\sin \theta}}
{\rd\over {\rd \phi}}\Bigr)(\sin^s \theta \,\n).  \eqno (A1.4b)
$$
The action of these operators on the spin-weighted spherical harmonics
can be found in Appendix A of [22]
and further details can be found in [23,26].

\bigskip

\noindent Appendix 2.  Properties of $Z^a{}_b$ \hfill\break

In Eq. (3.2e), we defined (a,b = 2,3)
$$
 Z^a{}_b\ :=\ \d^a{}_b \- f^a{}_b,  \eqno (A2.1a)
$$
$$
f^a{}_b \ :=\ { {\S_{\3}{}^a A^{\3}{}_b}\over { \S_{\3}{}^c A^{\3}{}_c}}
\eqno (A2.1b)
$$
It is easy to see that the determinent of $f^a{}_b$ is zero and there are
two eigenvalues (1,0) with the
respective eigenvectors $  \S_{\3}{}^a \and \e^{ab}A^{\3}_b$ and eigenforms
$ A^{\3}{}_a  \and
\e_{ab} \S_{\3}{}^b $ where $\e_{ab}$ is the two-dimensional Levi-Civita
tensor.    The same is true of
$ Z^a{}_b $ with the eigenvalues (0,1).  Therefore,
$ Z^a{}_b$  is a projection operator. In the $1/r$ expansion of the variables,
 the zeroth order term will
select eigenforms and eigenvectors of spinweight $ -1 \and +1$,
respectively.  More specifically, we have
$$
{}^0\!Z\= {\Io 2}\pmatrix {1& -i\sin \theta \cr {i/ \sin \theta} & 1\cr}
\eqno (A2.2)
$$
so that
$$
\eqalign{
{}^0\!Z^a{}_b s^b\= &a_a {}^0\!Z^a{}_b \= 0 \cr
{}^0\!Z^a{}_b t^b\= t^a,\quad &b_a {}^0\!Z^a{}_b \= b_b \cr
} \eqno (A2.3)
$$
The vectors and covectors $s^a, t^a, a_a, \and b_a$ are the spin-weighted
basis vectors and covectors
introduced in Appendix 1.
\medskip
\noindent Appendix 3.  The Solution for $V^a$.

We make no restrictions on $\S_{\3}{}^2$ or
$A^{\3}{}_2$ except that they should start with their Minkowski space
values at null infinity and have a $1/r$ expansion in the vicinity of
null infinity.  In addition there are the coordinate and tetrad conditions
which are described at the beginning of Section 3.  We could proceed as in
Section 4 specifying $\Sc^2 \and \Ac_2$, but equivalently we can write
$$
\eqalign{
\S_{\3}{}^a\= &-ir\sin\theta\{s^a\pl{\Io r}{}^1\!\S^a \pl{\Io {r^2}}{}^2\!\S^a\}
\cr
A^{\3}{}_a\= &-\{a_a\pl {\Io r}{}^1\!A_a\pl {\Io {r^2}}{}^2\!A_a \} \cr
}\eqno (A3.1)
$$
The restrictions which are a result of the equations $\G_1, \G_2, \and \H_1$
lead to the same result.  $\G_2$ tells us that $\Ac_1=0$, then $\G_1 \and \H_1$
give us
$$
\eqalign{
\S_{\3}{}^a A^{\3}{}_a &\= ir\sin\theta \cr
A^{\3}{}_{a,1}\S_{\3}{}^a &\= 0. \cr
}\eqno (A3.2)
$$
Substituting (A3.1) into these equations yields the result
$$
\eqalign{
\S_{\3}{}^a &\= -ir\sin\theta\{s^a\pl {\Io r}{}^1\!\S t^a \pl
	{\Io {r^2}}({}^2\!\S t^a- \h {}^1\!A{}^1\!\S s^a \} \cr
A^{\3}{}_a &\= -\{a_a\pl {\Io r}{}^1\!\A b^a \pl
	{\Io {r^2}}({}^2\!\A b^a- \h {}^1\!A{}^1\!\S a^a \} \cr
}\eqno (A3.3)
$$
{}From this point on, one just puts this result into the succeeding equations
and proceeds as before looking for the solutions.  Now some logarithmic
terms appear, but below the first couple of terms.  However, the problem arises
in the equations for $V^a,
\chi^a=0$, which do not depend on these solutions.  We have $(V^a:={\ut N}v^a)$:
$$
(V^a\Sa^1)_{,1} \- {\Io r}Z^a{}_b (V^b\Sa^1) \= 0. \eqno (A3.4)
$$
$$
\eqalign{
Z^a{}_b\ &:=\ \d^a{}_b \- {\Sc^a\Ac_b\over \Sc^c\Ac_c} \cr
	&:=\ t^ab_b - {\Io r}\bigl[s^ab_b{}^1\!A\pl t^ab_b{}^1\!\S\bigr]
	-{\Io r^2}\bigl[(s^aa_b+t^ab_b){}^1\!A{}^1\!\S + {}^2\!\S\bigr].\cr
}
$$
Write
$$
V^a\Sa^1\= -ir\sin\theta\bigl[{}^0\!\V^a+{{}^1\!\V^a\over r}+
{{}^2\!\V^a\over r^2}\bigr] \eqno (A3.5)
$$
and substitute into (A3.4).  We obtain
$$
\eqalign{
{}^0\!\V^a\- {{}^2\!\V^a\over r^2}\= &
\bigl[t^ab_b - {\Io r}\bigl(s^ab_b{}^1\!A\pl t^ab_b{}^1\!\S\bigr)
	-{\Io r^2}\bigl[(s^aa_b+t^ab_b){}^1\!A{}^1\!\S + {}^2\!\S\bigr]\cr
	&\qquad \times \bigl[{}^0\!\V^a+{{}^1\!\V^a\over r}+
{{}^2\!\V^a\over r^2}\bigr] \cr
}\eqno (A3.6)
$$
which gives the relations
$$
{}^0\!\V^a={}^0\!\V^bb_b\ t^a, \qquad ({}^1\!\V^a-{}^0\!\V^ba_b{}^1\!\S)t^a
	- {}^0\!\V^bb_b {}^1\!\A\ s^a\=0.
$$
Thus,
$$
{}^0\V^a a_a\={}^1\!\V^bb_b \= ({}^0\V^a b_a) {}^1\!A = 0. \eqno (3.7)
$$
The other expansion terms are given in terms of known quantities except for
${}^0\!\V^a b_a \and {}^1\!\V^a a_a$ which remains arbitrary.   In order that
$v^a= {\bar \Sc}^a$ when the reality conditions are imposed, we necessarily
choose ${}^1\!A = 0$ as noted in Section 5.  Then we have
$$
\eqalign{
V^a\= &-{\Io r}\bigl[({}^0\!\V +{\Io {2r^2}}{}^1\!\V{}^1\!\S\bigr) +
	({\Io r}{}^1\!\V+{\Io {2r^3}}{}^0\!\V{}^3\!A )], \cr
\Ac_a\= &-a_a\pl {{}^2\!A\over r^2}b_a +{{}^3\!A \over r^3}b_a.  \cr
} \eqno (A3.8)
$$

\vfil
\eject
\noindent References.

\item {1.}  Ashtekar, A. (1986) {\it Phys. Rev. Lett.}{\bf 57}, 2244.
\item {2.}  Ashtekar, A. (1987) {\it Phys. Rev. D} {\bf 36}, 1587.
\item {3.}  Ashtekar, A. (with invited contributions) (1988) {\it New
Perspectives in Canical Gravity} (Naples, Bibliopolis).
\item {4.}  Ashtekar, A. (1991) {\it Lectures on Non-perturbative
Canonical Gravity (Notes prepared in collaboration with R. Tate)}
(Singapore: World Scientific).
\item {5.}  Goldberg, J.N., Robinson, D.C., and Soteriou, C. (1992)
{\it Class. Quantum Grav.}{\bf  9} 1309.
\item {6.}  Goldberg, J.N. (1985) {\it Found. Phys.} {\bf 15}, 439.
\item {7.}  Torre, C. (1986) {\it Class. Quantum Grav.}{\bf  3}, 773.
\item {8.} Dirac, P.A.M. (1958) {\it Proc. R. Soc.} {\bf 246}, 333.
\item {9.} Dirac, P.A.M. (1959) {\it Phys. Rev.} {\bf 114}, 924.
\item {10.}  Bergmann, P.G. and Komar, A. (1960) {\it Phys. Rev. Lett.}
{\bf 4}, 432.
\item {11.}  Bondi, H., Van Der Burg, M., and Metzner, A. (1962) {\it
Proc. R. Soc. A} {\bf 269}, 21.
\item {12.}  Sachs, R. (1962) {\it Proc. R. Soc. A} {\bf  270}, 103.
\item {13.}  Newman, E.T. and Penrose, R. (1962) {\it  J. Math.
Phys.} {\bf  3}, 566.
\item {14.}  Newman, E.T. and Unti, T. (1962) {\it J. Math. Phys.}
{\bf  3}, 891.
\item {15.}  Chru\'ciel, P.T., MacCallum, M.A.H., and Singleton,
D.B. (1994) {\it Phil. Trans. of the R. Soc. London}, to appear.
\item {16.}  S. Novak and J.N. Goldberg (1981) {\it Gen. Rel. Grav.}
{\bf 13}, 79.
\item {17.} S. Novak and J.N. Goldberg (1982) {\it Gen. Rel. Grav.}
{\bf 14}, 655.
\item {18.}  Jacobson, T. and Smolin, L (1987) {\it Phys. Lett.}
 {\bf  196 B}, 39.
\item {19.}  Samuel, J. (1987) {\it Pramana J. Phys.} {\bf 28}, L429.
\item {20.}  Sachs, R. (1962) {\it J. Math. Phys.}{\bf 3}, 908.
\item {21.}  L. Tamburino and J. Winicour, (1966) {\it Phys. Rev.}
{\bf  150}, 1039.
\item {22.}  Glass, E.N. and Goldberg, J.N. (1970) {\it J. Math. Phys.}
{\bf 11}, 3400.
\item {23.}  Penrose, R. and Rindler, W. (1984) {\it Spinors and
Space-time} (Cambridge University Press, Cambridge), vol. 1.
\item {24.}  Winicour, J. (1985) {\it Found. Phys.} {\bf 15}, 605.
\item {25.} Dray, T. (1985) {\it Class. Quantum Grav.} {\bf 2}, L7.
\item {26.}  Newman, E.T. and Penrose, R. (1966) {\it J. Math. Phys.}
{\bf 7}, 863.

\end